\documentclass{an}
\usepackage{graphicx}
\usepackage{times}
\def\ergscm2{ \,{\rm erg}\,{\rm s}^{-1}\,{\rm cm}^{-2}\,}
\def\deg2{ \,{\rm deg}^{2}\,}

\Volume{01}              %
\Year{2002}              %
\Month{10}               %
\Pagespan{223}{227}      %

\begin{document}
\lhead[\thepage]{A.N. Author: Title}
\rhead[Astron. Nachr./AN~{\bf XXX} (200X) X]{\thepage}
\headnote{Astron. Nachr./AN {\bf 32X} (200X) X, XXX--XXX}

\title{DUET - The Dark Universe Exploration Telescope}

\author{K. Jahoda for the DUET collaboration}
\institute{
Laboratory for High Energy Astrophysics, Goddard Space Flight Center, Greenbelt MD 20771, USA
}

\date{Received {\it date will be inserted by the editor}; 
accepted {\it date will be inserted by the editor}} 

\abstract
{
DUET, a proposed NASA Midex class mission, will produce a catalog
of $\sim 20,000$ X-ray selected clusters and groups of galaxies suitable
for detailed cosmological investigations.  We present the 
cosmological
constraints achievable by DUET, the functional requirements of the survey,
and briefly describe an instrument/mission 
design capable of achieving these requirements.  The advantages of an X-ray survey
for constraining the nature and amount of dark energy and dark matter
are emphasized.  Constraints on cosmological parameters are precise
and complementary to other approaches.  The density of dark energy, 
dark matter, and neutrinos will be measured with a statistical uncertainty of\
0.5\%, 0.5\%, and 0.2\% of the
critical density and the dark energy equation of state can be measured to $\pm 0.2$
if dark energy and dark matter contribute
$\sim 2/3$ and $\sim 1/3$ of the critical density as currently believed.
DUET is a multi-national collaboration led by GSFC with
substantial contributions from groups in the US, Europe, and Japan.
}

\correspondence{keith.jahoda@gsfc.nasa.gov}

\maketitle

\section{Introduction}
The Dark Universe Exploration Telescope (DUET) is a MIDEX class concept that
would study the distribution and nature of dark energy and dark matter using
two large and deep X-ray selected samples of clusters of galaxies.  A "wide" sample
would reach limiting 0.5 - 2 keV fluxes of $5 \times 10^{-14} \ergscm2$ over the
$10^4 \deg2$ region of the Sloan Digital Sky Survey (SDSS).  A "deep" sample would
reach a limiting flux 10 times fainter over $100 \deg2$ in a southern hemisphere region that
overlaps a deep Sunyaev-Zeldovich (SZ) survey.
These catalogs can be obtained within a three year mission.  The wide sample will
contain 15-20,000 clusters or groups of galaxies;  the deeper survey 
will provide significant constraints on cluster evolution.
Together the two samples will provide powerful probes of 
cosmological models, precise estimates of fundamental cosmological parameters,
and clues to the nature of Dark Energy and Dark Matter.  The expected constraints are
highly complementary and provide similar or better statistical precision as
those expected from the next generation of Cosmic Microwave Background
experiments and the proposed SuperNova Acceleration Probe (SNAP).

DUET was proposed to the NASA MIDEX competition in 2001 by an international collaboration
led by the Goddard Space Flight Center (GSFC) with flight hadware, ground system, and
optical follow-up contributions 
from GSFC, the Massachusetts Institute of Technology, the European Space Agency, the Italian (ASI)
 and Japanese (ISAS) Space Agencies, and
the Universities of Illinois, Chicago, and Hawaii.  The Science Team includes members from
these institutions and Berkeley, Cal Tech, Princeton, ESTEC, Leicester, MPE, Brera,
Saclay, Santander, TiTech, and Trieste.

\section{From Clusters to Cosmology}
Current cosmological theories make precise predictions about the "halo mass function"
as a function of cosmological parameters (i.e. the Hubble constant $H_o$, the matter
density $\Omega_M$, the dark energy density $\Omega_E$, the fluctuation amplitude $\sigma_8$)
and redshift.  The space density and distribution of the most massive objects, which are
observationally identified with clusters of galaxies, are
very sensitive to cosmological
parameters.  Thus, a large catalog of clusters with a well defined and uniform
selection function and a robust observable-mass relationship
provides an excellent probe of cosmological parameters (\cite{HMH01}; \cite{LSM02}).

\begin{figure}
\resizebox{\hsize}{!}
{\includegraphics[angle=-90,width=5cm]{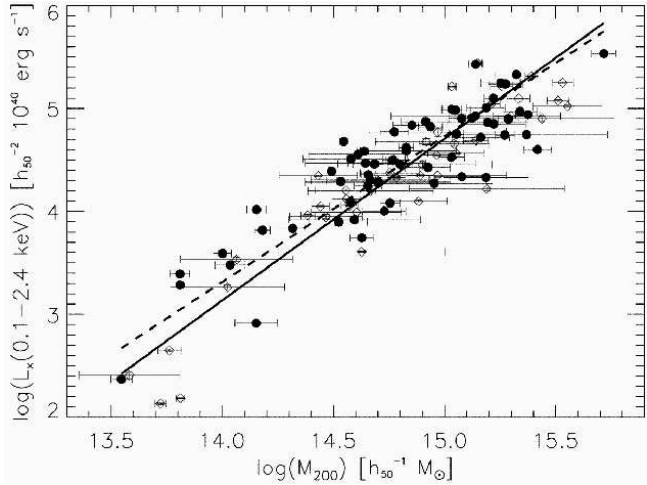}}
\caption{The Luminosity-Mass relationship for a large sample of clusters of galaxies
observed by ROSAT and ASCA from \cite{RB02}.  The scatter is 60\% in mass at a fixed
luminosity, which dominates statistical errors in estimating the mass.}
\label{lm_rel}
\end{figure}

The basic observational data provided by DUET is a catalog of about 20,000 clusters described
by spatial coordinates, angular size,  and X-ray flux;  for the brightest $\sim 1000$ objects, DUET
will also measure temperatures.
Correlation with
optical observations, either the SDSS or dedicated follow-up, provides redshift information
allowing a reliable determination of X-ray luminosity.  The mass of each cluster can be
estimated from the $L-M$ relationship (fig. \ref{lm_rel}) which
is well defined over a broad range of cluster masses and luminosities
(\cite{RB02}).  The observed distribution of masses can
be compared with cosmological predictions;  there are numerous checks on the mass scale
including the DUET temeperatures for the brightest objects, comparison with detailed
$\it Chandra$ and XMM-Newton observations, correlations with masses derived from
SZ, strong and weak lensing, and velocity dispersion surveys.

X-rays are an excellent cluster tracer since
most of the
baryonic (i.e. potentially visible matter) resides in a hot diffuse and X-ray emitting
gas which permeates
the potential well of the cluster.  
The temperature of the gas
is primarily determined by gravitational heating.  Although the $L-M$ relationship 
is steeper than predicted from gravitational heating alone,
it is well defined and predictive over several orders of magnitude in both
quantities.  Several pre-heating models are capable of explaining the observed
$L-M$ relationship, and the complete physics based
description is likely to be well
understood in detail prior to the DUET mission.

Clusters are bright objects in the X-ray sky, and at high galactic latitudes are the only common
extended sources.  They are easy to identify from X-ray data with positions adequate
to provide optical identifications and redshifts.

\section{Constraints}
\begin{figure}
\resizebox{\hsize}{!}
{\includegraphics[angle=0,width=5cm]{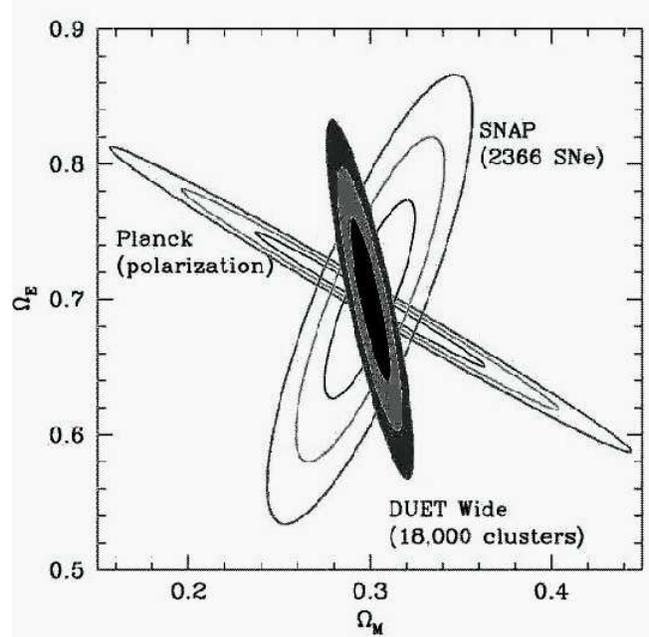}}
\caption{DUET constraints on the matter and dark energy density (marginalized over
$w$ and $H_o$), compared with
measurements from the CMB and the proposed SNAP survey.  Note that the DUET
constraints have similar precision while being highly complementary to the other
constraints.}
\label{omegaML}
\end{figure}

Precise cosmological inferences can be made by statistical analyses
of the DUET three dimensional map of the distribution of clusters of galaxies.
Figures \ref{m_nu} through \ref{omegaML} show examples.  These figures are based on
simulations using a fiducial cosmology
\begin{footnote}
{The fiducial cosmology for the
simulations presented here assumes a total matter density $\Omega_M = 0.3$, a dark
energy density $\Omega_E = 0.7$, 
a Hubble constant of $100 h {\rm km}\,{\rm s}^{-1}\,{\rm Mpc}^{-1}$
with $h = 0.72$, a baryon density $\Omega_b h^2 = 0.02$, a dark energy equation of state
parameter $w = -1$, a power spectrum normalization $\sigma_8 = 1.0$,
a fluctuation spectrum index $n_s = 1$, and a neutrino density
$\Omega_\nu = 0$.}
\end{footnote}
and demonstrate the precision with
which parameters can be determined.  The error contours are small, and they are
significantly tilted with respect to the contours that will be determined by other planned or
proposed experiments such as Planck and SNAP (fig. \ref{omegaML}).
This means that the combined experiments will either determine with great precision the
values of many cosmological parameters contained in currently popular cosmological models,
or more excitingly, 
reveal discrepancies with CMB/SNAP, and thus  point to a need for a
fundamental reassesment of cosmological models

\begin{figure}
\resizebox{\hsize}{!}
{\includegraphics[angle=-90,width=5cm]{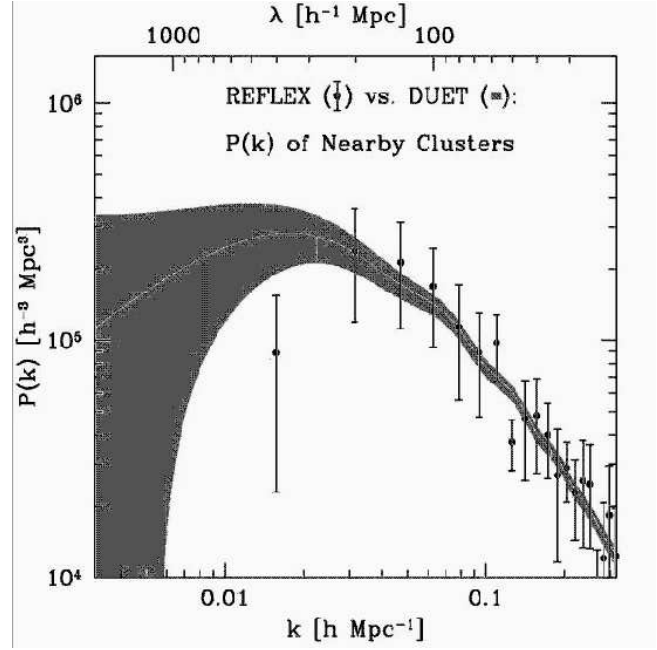}}
\caption{Comparison of the REFLEX power spectrum with the DUET power spectrum for
clusters with redshift less than 0.25.  DUET will measure independent power spectra, of
similar statististical precision, for the redshift range 0.25 to 0.45 and for redshifts
above 0.45.}
\label{ps}
\end{figure}

Two powerful statistical tests are the power spectrum, $P(k)$, and
the redshift distribution, $dN/dz$.  
The state of the art in measurement of the power spectrum is derived from the 
REFLEX catalog (\cite{B01}) with almost
500 objects in the southern sky.  DUET will provide 40 times more objects
in a comparable solid angle.
Figure \ref{ps} compares the power spectrum measured from the REFLEX survey with that
expected from DUET clusters with redshift below $0.25$ (i.e. about one third of the DUET
survey selected in the redshift range sampled by REFLEX.)
Interpretation of both $dN/dz$ and $P(k)$ benefits enormously from a uniform and
well defined survey.  Although XMM-Newton will discover a substantial number of clusters
serendipitously (\cite{R01}), and provide information complimentary to DUET,
the XMM serendipitous catalog will be constructed from numerous pencil
beam observations of various depths, will generally
probe fainter fluxes (and lower masses), will often observe directions not
covered by the SDSS, and does not have a dedicated optical follow-up program
to provide redshifts for the balance of the clusters.

\subsection{Power Spectrum}
The power spectrum measures, as a function of scale, the lumpiness of the universe (\cite{BG01})
and provides a more complete description than $\sigma_8$ (
the
variance of the density field on 8 Mpc scales) which is often used to normalize the power
spectrum and which can be related to P($0.2 \, h \, {\rm Mpc}^{-1}$) (\cite{H02}).

\begin{figure}
\resizebox{\hsize}{!}
{\includegraphics[angle=0,width=5cm]{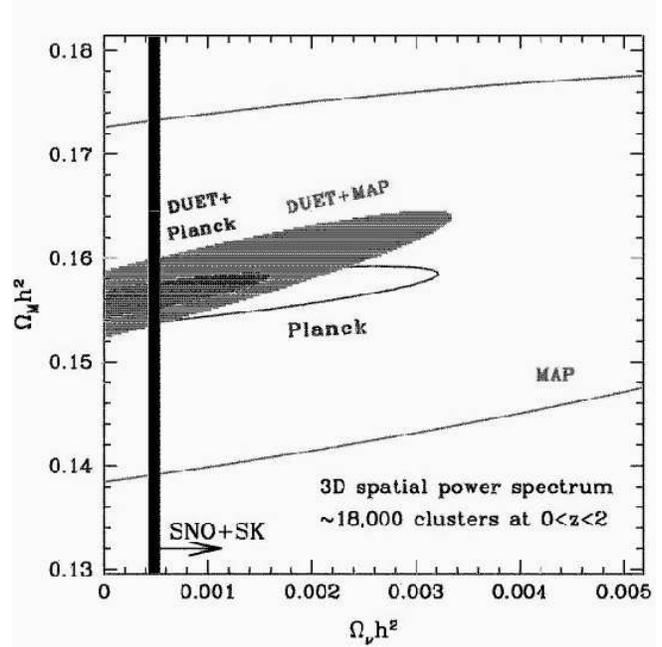}}
\caption{DUET constraints on the energy density of neutrinos.  The shape
of the power spectrum depends on the fraction of clustered and non-clustered
matter.  Neutrinos are not expected to cluster on the scales of clusters of
galaxies;  DUET can measure the presence of non clustered dark matter at
a few tenths of a percent of the critical density.}
\label{m_nu}
\end{figure}

Even in the context of cosmological models dominated by dark matter and dark energy,
the
overall shape of the power spectrum is quite sensitive to small amounts of baryons or
neutrinos, whose clustering properties are different than for cold dark matter alone.
For baryons, the effects arise from numerous non-gravitational interactions
including acoustic oscillations (which produce the beautiful peaks in the microwave
background power spectra) and Compton drag (see \cite{EH98} for a detailed description).
A neutrino (or Hot Dark Matter) contribution to the energy density of the universe would
suppress structure at scales below $ k > \approx 0.05 h {\rm Mpc}^{-1}$ (\cite{E02}).
Figure \ref{m_nu} shows that DUET could provide an interesting upper limit to the
energy density in neutrinos (and thus neutrino mass).  The lower limit in the figure
comes from Super-Kamiokande results (\cite{A01}).   

\subsection{Redshift Distribution}
\begin{figure}
\resizebox{\hsize}{!}
{\includegraphics[angle=0,width=5cm]{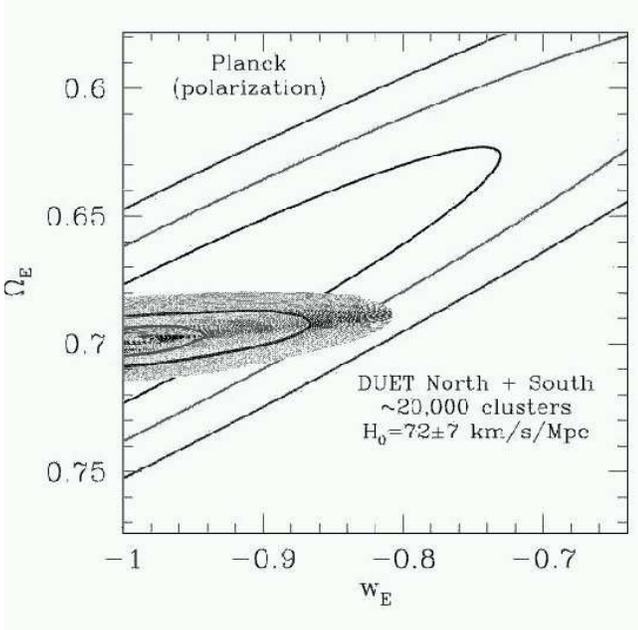}}
\caption{DUET constraints on the Dark energy density and the dark energy equation of
state parameter $w$ assuming a flat universe;  the contours are marginalized over
$\sigma_8$ and $H_o$.}
\label{omegaE_w}
\end{figure}

For a non-evolving tracer, the redshift distribution $dN/dz$ is 
a probe of cosmological volume (and thus
of the energy density of dark energy and dark matter). 
The ideal tracer has a constant (or calculable) volume density
and a well known selection function.  X-ray selected clusters of galaxies
satisfy these conditions.
While clusters of galaxies do evolve, 
their growth is dominated by gravitational collapse.
This makes the volume density directly calculable without including the difficult
physics involved in galaxy formation.  Recent numerical simulations
demonstrate the existence of a universal mass function, describable in terms of
cosmological parameters and the cluster growth function for a wide range of cosmological
models (\cite{J01}).  Evolution of the cluster volume density provides tight constraints
as discussed in detail 
detail by \cite{HMH01} and demonstrated in figures \ref{omegaE_w} and \ref{omegaML}.  

\section{DUET Pathfinder}
The ROSAT North Ecliptic Pole (NEP) survey (\cite{H01}) has measured the X-ray flux distribution of
clusters, and demonstrates in a model independent way that DUET will detect 15-20,000 clusters
in the SDSS region.
The NEP survey covered $81 \deg2$ at the north ecliptic pole
and reached a limiting sensitivity of a few  $10^{-14} \ergscm2$.  An eight year optical
follow-up effort identified all but 2 of the 445 sources.  The survey includes 64 clusters
with median and maximum redshift of 0.205 and 0.811.  

The complete identification of NEP sources allows quantitative statements about the 
optical magnitudes of the sources.  Mullis (2001) finds that the median plus 1 $\sigma$
B magnitude for the AGN is 20.7 (i.e. 190 of the 218 AGN are brighter than this), 
and all of the AGN have B magnitudes less than 22.  Stars also have magnitudes as faint
as 22, but the mean plus $1 \sigma$ point (137 of 152) is at B = 17.2.
The details
of the DUET stellar content are expected to be latitude dependent.
The depth  of the SDSS photometric survey is well suited to
identify the vast majority of the X-ray point sources in the
DUET catalog, in addition to providing redshifts for the clusters.

\section{Requirements of an X-ray Survey}
The observational requirements of DUET collectively dictate the size and quality of the
X-ray optics and focal plane.  Required features include
\begin{itemize}
\item{The ability to distinguish extent from X-ray data alone, in order to separate clusters
of galaxies from other the high latitude sources which are pointlike}
\item{Small enough error circles to provide unambiguous identifications}
\item{Small systematic mass errors}
\item{Sufficient grasp to obtain a large catalog within the scope of a MIDEX mission}
\end{itemize}

Requiring 50 photons for threshold
objects, in combination with the proposed optics, satisfies the first two requirements.
The tight observed $L-M$ relationship and the ability to calibrate DUET masses against
detailed $\it Chandra$ and XMM-Newton observations, and against masses derived from
SZ, strong and weak lensing, and velocity distpersions satisfies the third.
The grasp available from the flight spare mirror for
the X-ray Multi-Mirror (XMM) mission can achieve these results in
a three year mission.

The XMM mirror has substantial effective area out to a radius of 30', albeit with degraded
angular resolution.  Weighted sums of ray tracing results allow the investigation of the
field of view averaged point spread function, which has a $90\%$ encircled energy radius of
46 arcsec, and a vignetting relative to the on-axis effective area of 0.3.  We have performed
simulations under the assumption that 50 photons are detected from a source with a $\beta$
model surface brightness profile (\cite{CF76}); we assume $\beta = 0.7$.  
The radial profile of the source photons (including
appropriately scaled contributions from instrumental and cosmic backgrounds) are compared
to the point spread function, and a Kolmogorov-Smirnov test is employed to determine
whether the source is extended.  
50 photons are sufficient to identify ($99\%$ confidence level) 70, 92, and 98\% of 
clusters with core radii of 20, 30, and 40 arcsec respectively.

The typical core radius for clusters is 220 kpc (\cite{V99}) and 
ROSAT samples suggest that core radii do not evolve (\cite{V98}), suggesting that
the typical cluster has a minimum angular core-radius of 30 arcsec for redshifts
near 1 in flat universes.  
Even if core-radii evolve, the XMM mirror should detect as extended the large majority of clusters,
with a calculable and calibratable incompleteness for the smaller systems.

\section{DUET Instrument Implementation}
The DUET concept takes advantage of the XMM flight spare mirror, i.e. an excellent
optic which already exists and which is well calibrated.  The
desired field of view is 1 degree (diameter).  The focal length is 7.5 m.
The field of view is achieved with a 4 by 4 array of CCD chips while the focal
length is accommodated by a folding optical bench.

The optical bench consists of two triangular truss sections constructed from flat
composite panels connected by a hinge and
latch system designed to be operated once, after a suitable orbit is achieved.
The instrument can be accommodated on a modified Ball Aerospace BCP 2000 spacecraft
bus and a Delta II launch vehicle.

The SDSS will provide redshifts for 85\% of the clusters.  A 
dedicated 3 year optical followup program will complete the DUET catalog.

\acknowledgements
I thank the organizers of the X-ray Surveys conference for a stimulating
conference.  I also thank my collaborators on the DUET proposal who have
contributed to the concepts and simulations presented in this summary.

\end{document}